\documentclass{article}
\usepackage{spconf,amsmath,graphicx,hyperref}
\usepackage{amssymb}
\usepackage{amsmath}
\usepackage{mathtools}
\usepackage{booktabs}
\usepackage{multirow}
\usepackage{newtxmath}
\usepackage{bm}

\title{Look, Listen and Segment: Towards Weakly Supervised Audio-visual Semantic Segmentation}
\name{Chengzhi Li \qquad Heyan Huang \qquad Ping Jian\sthanks{Corresponding author.} \qquad Yanghao Zhou }
\address{School of Computer Science\\
  Beijing Institute of Technology\\
	Beijing, China}
\begin{document}
\topmargin=0mm
\maketitle
\begin{abstract}

Audio-Visual Semantic Segmentation (AVSS) aligns audio and video at the pixel level but requires costly per-frame annotations. We introduce \textbf{W}eakly \textbf{S}upervised \textbf{A}udio-\textbf{V}isual \textbf{S}emantic \textbf{S}egmentation (WSAVSS), which uses only video-level labels to generate per-frame semantic masks of sounding objects. We decompose WSAVSS into looking, listening, and segmentation, and propose \textbf{P}rogressive \textbf{C}ross-modal \textbf{A}lignment for \textbf{S}emantics (PCAS) with two modules: \textit{Looking-before-Listening} and \textit{Listening-before-Segmentation}. PCAS builds a classification task to train the audio-visual encoder using video labels, injects visual semantic prompts to enhance frame-level audio understanding, and then applies progressive contrastive alignment to map audio categories to image regions without mask annotations. Experiments show PCAS achieves state-of-the-art performance among weakly supervised methods on AVS and remains competitive with fully supervised baselines on AVSS, validating its effectiveness.

\end{abstract}

\begin{keywords}
Audio-Visual Semantic Segmentation,  Weakly Supervised Learning,  Contrastive Learning
\end{keywords}

\section{Introduction}
\label{sec1}
Audio-Visual Source Segmentation (AVS) targets fine-grained grounding of sounding regions~\cite{wei2022learningaudiovisualcontextreview}. Its semantic extension, Audio-Visual Semantic Segmentation (AVSS), further requires class-specific masks for multiple sounding objects in multi-source scenes~\cite{zhou2023avss}. AVSS demands more costly semantic annotations than AVS. To reduce this burden, we introduce \textbf{W}eakly \textbf{S}upervised \textbf{A}udio-\textbf{V}isual \textbf{S}emantic \textbf{S}egmentation (WSAVSS), which uses only video-level labels to generate per-frame semantic masks of sound sources. WSAVSS exploits richer cross-modal structure without dense annotations but faces two primary challenges:

\textbf{Frame-Precise Auditory Events Localization.} Classic fully supervised AVSS trains the whole model (including the audio encoder) with per-frame per-pixel labels, uniformly processing all audio across frames~\cite{zhou2022avs}. Without such dense labels, it cannot localize sounding objects frame by frame, hindering frame-wise semantic segmentation. We propose the \textit{Looking-before-Listening} module: the model first “looks” to obtain a stable visual prior, then “listens” to strengthen per-frame audio understanding. We sequentially insert frame semantic features as prompt tokens into the audio token sequence, enabling frame-precise auditory understanding without frame-level segmentation annotations. We term this \textbf{T}emporal \textbf{V}isual \textbf{P}rompting (TVP).

\textbf{Pixel-level Cross-modal Semantic Mapping.} Numerous fully supervised approaches have been proposed to achieve fine-grained semantic alignment~\cite{zhou2022avs, zhou2023avss,  tommzhou, Gao_Chen_Chen_Wang_Lu_2024, yang2023cooperation}. However, none of these methods explore how to achieve fine-grained semantic alignment without segmentation annotations. Inspired by ToCo~\cite{toco}, we introduce cross-modal token-wise contrastive learning for fine-grained alignment, with contrast labels are derived from cross-modal similarities produced by other instance-level contrast module, enabling progressive coarse-to-fine alignment. This ensures the model conducts fine-grained semantic segmentation only after fully understanding the audio (the \textit{Listening-before-Segmentation} module), thereby addressing the challenge of cross-modal semantic mapping.

Integrating the \textit{Looking-before-Listening} and \textit{Listening-before-Segmentation} modules, we present \textbf{P}rogressive \textbf{C}ross-modal \textbf{A}lignment for \textbf{S}emantics (PCAS), and extensive experiments verify its effectiveness. Our contributions are summarized as follows:

\begin{itemize}

    \item We introduce WSAVSS, which offers a challenging yet valuable task for advancing weakly supervised fine-grained audio-visual understanding.
    \item We propose TVP, which uses raw visual cues as prompts to guide frame-precise audio understanding under weak supervision.
    \item We design a progressive contrastive alignment framework combining instance-wise and token-wise contrast to ensure solid audio understanding before segmentation for WSAVSS.

\end{itemize}

\section{RELATED WORK}
\label{sec:related_work}
The related work is summarized in two aspects: 

\textbf{Audio-Visual Segmentation. }
In 2022, Zhou et al. introduced AVS, which predicts pixel-level binary masks of sounding objects to improve visual detail understanding~\cite{zhou2022avs}. AVSS extends this by producing class-specific masks for multiple sounding sources in multi-object scenes~\cite{zhou2023avss}, but this finer supervision further increases annotation cost.

\textbf{Weakly Supervised Semantic Segmentation. }
Weakly Supervised Semantic Segmentation (WSSS) trains segmentation models using weaker signals than pixel-level masks (e.g. class labels)~\cite{wsss_Survey}. With ViT’s rise~\cite{vit}, ViT-based methods have been explored for WSSS~\cite{toco}. Mo et al. (2023) introduced the first multimodal WSSS task, WSAVS~\cite{mo2024weakly}, but AVS remains oversimplified and limited in significance.

Therefore, we propose WSAVSS, which explores weak supervision in the more valuable AVSS task. To our knowledge, this is the first work to discuss WSAVSS.

\begin{figure}[tbp]
	\centering
	\includegraphics[width=\linewidth]{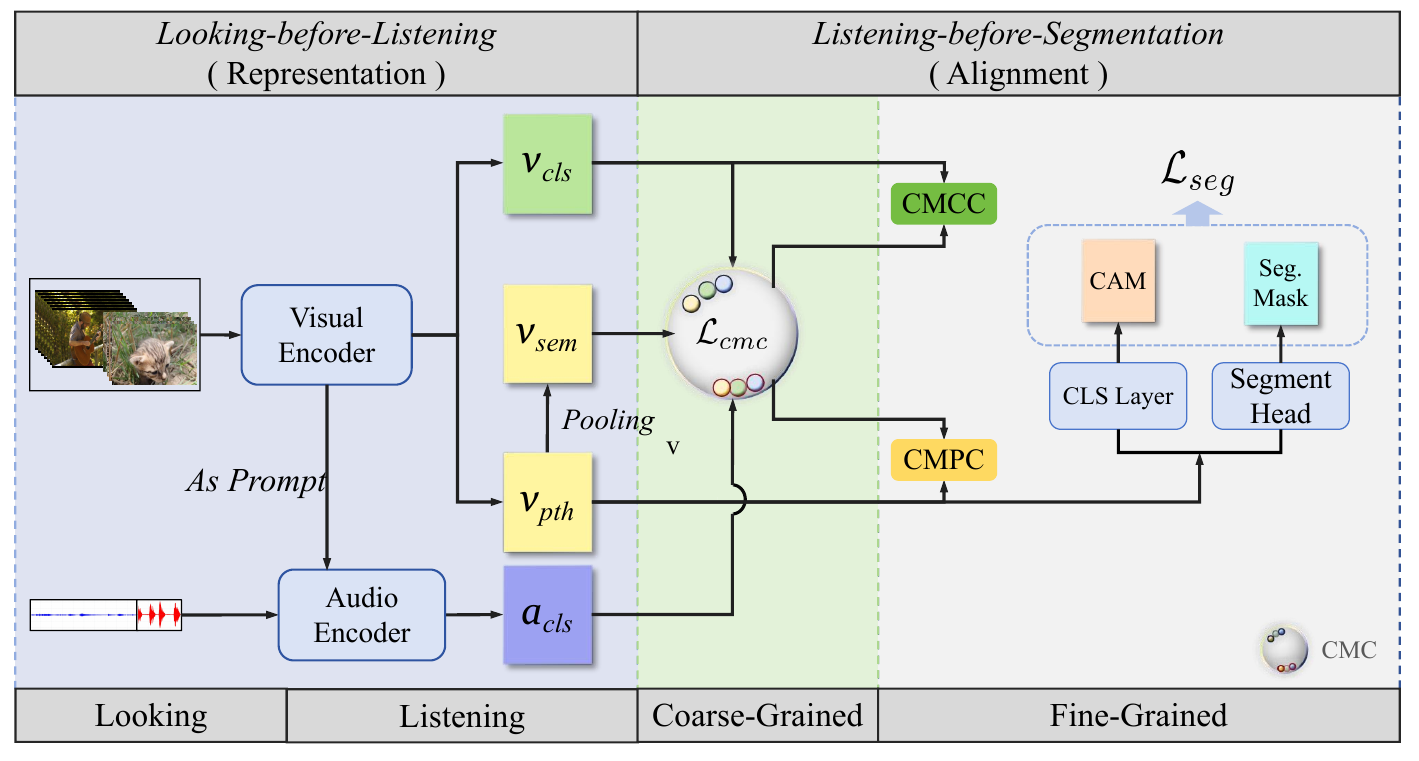}
	\caption{Overview of PCAS: The audio encoder processes audio frame-wise with visual prompts from the visual encoder. CMC  aligns global semantics; CMPC and CMCC refine fine-grained alignment via cross-modal similarity. The decoder then outputs masks using CAM-based pseudo-labels.
	}  \label{fig:framework}
\end{figure} 

\section{Our Method}
Motivated by basic human perception, we decompose WS-AVSS into three steps—looking, listening, and segmentation, highlighting the progressive order \textit{Looking-before-Listening} and \textit{Listening-before-Segmentation} (Figure~\ref{fig:framework}).

\subsection{Looking-before-Listening}
Human perceptions of images are often more vivid and lasting than those of audio, so viewing the image before listening allows for more accurate comprehension of the audio. Reversing this perception order does not produce the same effect.
Motivated by this phenomenon, we introduce \textbf{T}emporal \textbf{V}isual \textbf{P}rompting (TVP), which enhances audio comprehension by providing complementary visual information. This module ensures the accuracy of semantic understanding under weakly supervised settings.

We use ViT~\cite{vit} as the visual encoder, trained with video-level class labels. At each corresponding time step, its output pooled patch token ${v}^t_{sem}$ is inserted into the audio token sequence as a semantic prompt, providing complementary visual guidance for that segment audio. Using this token sequences training audio encoder to classify the audio segment and output pooled audio patch token ${a}^t_{sem}$.

\begin{figure}[tbp]
	\centering
	\includegraphics[width=0.9\linewidth]{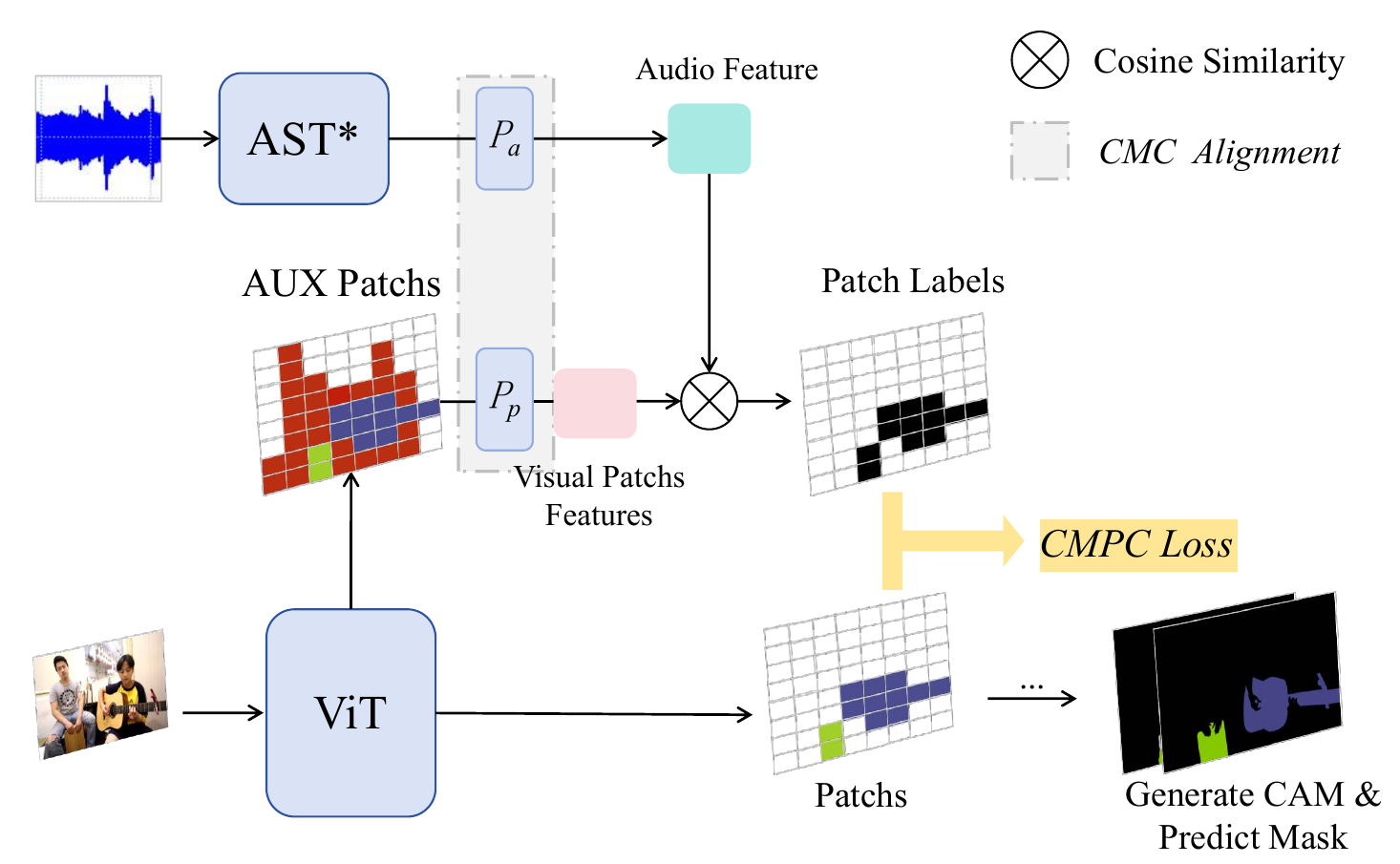}
	\caption{CMPC loss computation. ViT, AST*, $P_a$, $P_p$ denote the visual encoder, audio encoder with visual prompt, and projection layers for audio semantic tokens and visual patch tokens. Patches, AUX Patches, Patch Labels are ViT patch outputs, intermediate auxiliary patches, and positive/negative token-wise contrast labels.}
	\label{fig:CMPC_method}
\end{figure}

\subsection{Listening-before-Segmentation}
\textquotesingle \textit{Listening-before-Segmentation}\textquotesingle{} refers to thoroughly comprehending the audio semantics before segmenting the sounding objects. Through cross-modal contrastive learning between instances, multimodal tokens are aligned to the same feature space, after which tokens are labeled based on their cross-modal similarity for token-wise contrastive learning. Progressive alignment ensures that the model can accurately grasp the details of the image without mask supervision.

\subsubsection{Instance-wise Contrastive Learning}

Leveraging video-level labels, we build a supervised contrastive module \textbf{C}ross-\textbf{M}odal \textbf{C}ontrast (CMC) for global audio–visual semantic alignment. It projects tokens into a shared semantic space (as Coarse-Grained Semantic Alignment in Figure~\ref{fig:framework}). The objective loss is as follows:

\begin{equation}\small
	\begin{aligned}
		\mathcal{L}_{cmc}=\sum_{\mathbf{S}\in \mathfrak{S}} & (\mathtt{CE}(\mathbf{S},\mathbf{L})+\mathtt{CE}(\mathbf{S}^\top,\mathbf{L})),
		\\
		\mathfrak{S}=                                       & \{
		\mathbf{S}_{asem\_vsem}, \mathbf{S}_{asem\_vcls}, \mathbf{S}_{vcls\_vsem}\}
	\end{aligned}
\end{equation}
where \(\mathbf{L}, {a}_{sem}, {v}_{sem}, {v}_{cls} \) represent the label consistency matrix, audio semantic token, video semantic token and video classification token, respectively.  $\mathbf{S}_{x\_y}$ means cosin similarity between $x$ token and $y$ token. $\text{CE}(\cdot)$ denotes the \textbf{C}ross \textbf{E}ntropy loss function.

\begin{table}[tbp]

	\centering
	\setlength{\tabcolsep}{2.5pt}
	\fontsize{9pt}{10pt}\selectfont
	\begin{tabular}{cccccccc}
		\toprule
				\multirow{2}{*}{Methods}    & \multirow{2}{*}{Backbone} & \multicolumn{2}{c}{AVS-S4} & \multicolumn{2}{c}{AVS-MS3} \\
& & F-score & mIoU & F-score & mIoU  \\
    	\midrule

				 AVS (ws)~\cite{zhou2022avs} &ResNet-50 & 24.99 & 12.63 & 15.72 & 8.76  \\
CAM~\cite{cam} &ResNet-50 & 27.88 & 19.26 & 19.83 & 12.65  \\
EZ-VSL~\cite{mo2022localizing} &ResNet-50 & 35.70 & 29.40 & 27.31 & 23.58  \\
 \( \text{C}^{2}\text{AM} \)~\cite{c2am} &ResNet-50 & 36.55 & 30.87 & 29.58 & 25.33  \\
WS-AVS~\cite{mo2024weakly} & ResNet-50 & 51.76 & 34.13 & 46.87 & 30.85  \\
				\midrule

PCAS (Ours)                        & ResNet-50          & \underline{68.5}                       & \underline{56.41}                       & \underline{51.7}        & \underline{45.76}  \\
		 
				 PCAS (Ours)  & ViT-base  & $\bm{74.2}$ & $\bm{60.50}$ & $\bm{60.0}$ &  $\bm{46.04}$ \\

										\bottomrule
		
	\end{tabular}
\caption{Performance comparison with other weakly supervised baseline methods on the AVS-S4 and AVS-MS3 subsets.}
\label{tab:contrast}
\end{table}
\subsubsection{Token-wise Contrastive Learning}

Using CMC-generated positive/negative labels, we introduce cross-modal token-level contrastive learning to highlight audio-relevant patches (as Fine-Grained Semantic Alignment in Figure~\ref{fig:framework}). Specifically, we design \textbf{C}ross-\textbf{M}odal \textbf{P}atch token \textbf{C}ontrast (CMPC) and \textbf{C}ross-\textbf{M}odal \textbf{C}lass token \textbf{C}ontrast (CMCC) for sounding object perception.

\textbf{CMPC }
We compute the similarity between each visual patch token \({v}_{pth}\) and the audio token \({a}_{sem}\), build a patch-wise cross-modal similarity matrix, then apply a threshold to assign positive/negative labels for token contrastive learning, enhancing discrimination between sounding and non-sounding patches. The CMPC loss is computed as follows:

\begin{equation} \small
  \begin{aligned}
    \mathcal{L}_{cmpc} & = \sum_{t \in T} \Bigg[ \frac{1}{{N}^+}\sum_{\mathbf{L}_i = \mathbf{L}_j}(1-\mathtt{CS}(\mathbf{v}^{t}_i,\mathbf{v}^{t}_j)) \\
                      & +\frac{1}{{N}^-}\sum_{\mathbf{L}_i \neq \mathbf{L}_j}\mathtt{CS}(\mathbf{v}^{t}_i,\mathbf{v}^{t}_j) \Bigg]
  \end{aligned}
  \label{eq_loss_ptc}
\end{equation}

where \(\mathtt{CS}(\cdot,\cdot)\) is cosine similarity; \(T\) the time-step set; \(\mathbf{v}^t_i,\mathbf{v}^t_j\) the \(i\)-th and \(j\)-th visual patch tokens at time \(t\); \(\mathbf{L}\) the contrast label matrix from cross-modal similarity; \(N^+,N^-\) the counts of positive and negative pairs. See Figure~\ref{fig:CMPC_method}.

\textbf{CMCC }
We contrast local crop classification tokens with the global classification token: crops containing sounding regions are positives, others negatives. This drives the global token to focus on sounding objects while suppressing background and silent regions. Since positives labels are sparse, we adopt InfoNCE~\cite{cpc} as the CMCC objective. The CMCC computation is:

\begin{equation}\small
  \mathcal{L}_{cmtc} = \frac{1}{N^+}\sum_{\mathbf{c^{+}}}\log{\frac{e^{(\mathbf{g}^\top\mathbf{c^{+}}/\tau)}}{e^{(\mathbf{g}^\top\mathbf{c^{+}}/\tau)}+\sum_{\mathbf{v^{-}}}{e^{(\mathbf{g}^\top\mathbf{c^{-}}/\tau)}+\epsilon}}},
  \label{eq_loss_ctc}
\end{equation}

where \(\mathbf{g}\) is the global class token (i.e. $v_{cls}$); \(\mathbf{c}^+\) and \(\mathbf{c}^-\) are positive and negative local crop class tokens; \(\tau\) is the temperature; \(N^+\) the number of positives; \(\epsilon\) a small constant.

\section{Experiments}
\subsection{Experimental Settings}

We resize frames to \(10\times3\times448\times448\) for the AVSS subset and \(5\times3\times224\times224\) for other datasets. Audio is cropped to 10\,s and converted to log Mel filter bank (fbank) features. Training uses Adam for 9 epochs with 2 warm-up epochs; module learning rates are 0.0012 and 0.0006. Dense CRF~\cite{krahenbuhl2011efficient} refines the masks. Following prior work~\cite{zhou2022avs,zhou2023avss,tommzhou,Gao_Chen_Chen_Wang_Lu_2024,yang2023cooperation}, we report mean IoU (mIoU) and F-score for comparison.

\subsection{Comparative Experiments}

As the first weakly supervised approach for AVSS, no direct baselines exist on AVS-Semantic. Thus we run two comparisons: (1) against WSAVS methods on AVS-S4 and AVS-MS3 to show our advantages in fine-grained (especially boundary) alignment; (2) against fully supervised methods on the challenging AVS-Semantic subset to explore the upper limit of our method.

\begin{table}[tbp]
  \centering
	\setlength{\tabcolsep}{2.5pt}
	\fontsize{9pt}{10pt}\selectfont

  \begin{tabular}{ccccccc}
    \toprule
    \multirow{2}{*}{\begin{tabular}[c]{@{}c@{}}Training \\ Setting\end{tabular}}  & \multirow{2}{*}{Methods}   & \multirow{2}{*}{Backbone} & \multicolumn{2}{c}{AVS-Semantic}  \\
      &    & & F-score & mIoU \\
    \midrule
    \multirow{5}{*}{\begin{tabular}[c]{@{}c@{}}Fully\\Supervised\end{tabular}} 
    & AVS~\cite{zhou2022avs} & PVT-v2 & 42.39 & 29.77  \\
    & CATR~\cite{CATR} & PVT-v2 & 38.5   & 32.8  \\
        & AVSegFormer~\cite{Gao_Chen_Chen_Wang_Lu_2024}& PVT-v2 & 42.8 & 37.31 \\
        & AVSAC~\cite{AVSAC} & PVT-v2 & 42.39 & 36.98  \\
	& AVS-Mamba~\cite{gong2025avsmambaexploringtemporalmultimodal} & PVT-v2 & 45.1 & 39.7 \\

    & COMBO~\cite{yang2023cooperation} & PVT-v2 & $\underline{46.1}$ & $\bf{42.1}$ \\

        \midrule
    \multirow{2}{*}{\begin{tabular}[c]{@{}c@{}}Weakly\\Supervised\end{tabular}} & PCAS (Ours) & PVT-v2 & 44.6 & 36.30        \\

            & PCAS (Ours) & ViT-base & $\bm{52.2}$ & $\underline{42.07}$     &  \\

                                                    \bottomrule
  \end{tabular}
  \caption{Performance comparison with classic fully supervised methods on the AVS-Semantic subsets.}
  \label{tab:contrast_avss}
\end{table}

Tables~\ref{tab:contrast} and~\ref{tab:contrast_avss} show clear gains over all weakly supervised baselines under the same backbone on AVS-S4 and AVS-MS3, and competitive results versus fully supervised methods. \textbf{These results validate the effectiveness of our approach.} With both ResNet-50 and ViT-base, our method surpasses existing weakly supervised methods on AVS-S4 and AVS-MS3. On AVS-Semantic, ViT-base is competitive with top fully supervised models, and PVT-v2 matches classic fully supervised performance. \textbf{This indicates robustness to backbone choice.}

\begin{table}[tbp]
	\setlength{\tabcolsep}{8pt}
	\fontsize{9pt}{10pt}\selectfont
	\centering
	\begin{tabular}{lccc}
		\toprule
		Ablation Setting   & AVS-S4       & AVS-MS3      & AVSS         \\
		\midrule
		AST               & 0.481        & 0.205        & 0.158        \\
		AST + TVP           & 0.880        & 0.293        & 0.627        \\
						\bottomrule
	\end{tabular}
	\caption{The ablation results of the \textit{Looking-before-Listening}.}
	\label{tab:ablation1}
\end{table}

\begin{table}[tbp]
	\setlength{\tabcolsep}{2pt}
	\fontsize{9pt}{10pt}\selectfont

	\centering
	\begin{tabular}{cccccccccc}
		\toprule
		\multirow{2}{*}{\begin{tabular}[c]{@{}c@{}}Ablation \\ Setting\end{tabular}} 
		& CMC          
		& \begin{tabular}[c]{@{}c@{}}CMPC\end{tabular} 
		& \begin{tabular}[c]{@{}c@{}} CMCC\end{tabular} 
						& \multicolumn{2}{c}{AVS-Semantic} 
		\\
 & & & 
  & F-score & mIoU 
  \\
		\midrule
		\begin{tabular}[c]{@{}c@{}} w/o CMC, CMCC \\ \&CMPC  \end{tabular} & & & 
				& 0.367 & 25.318 
		\\
		\midrule
		w/o CMC\&CMPC& & & \checkmark 
																				 & 0.453 & 35.386 
										 \\
		w/o CMC\&CMCC& & \checkmark & 
																				& \underline{0.477} & \underline{37.485} 
										 \\
		w/o CMPC\&CMCC& \checkmark & & 
																				& 0.458 & 35.623
										 \\
		\midrule
		w/o CMPC& \checkmark & & \checkmark 
														  & 0.484 & 38.158
							   \\
		w/o CMCC& \checkmark & \checkmark & 
														  & \underline{0.500} & \underline{39.506}
							  \\
		w/o CMC & & \checkmark & \checkmark 
														  & 0.480 & 37.785
							  \\
		\midrule
		      PCAS & \checkmark & \checkmark & \checkmark 
						  & $\bm{0.522}$ & $\bm{42.074}$ 
			  \\
		\bottomrule
	\end{tabular}
	\caption{The ablation results of the \textit{Listening-before-Seg\\mentation} module.}
	\label{tab:ablation2}
\end{table}

\subsection{Ablation Study}

\textbf{Looking-before-Listening.} We employ a frame-by-frame audio classification task as the ablation study task. As shown in Table~\ref{tab:ablation1}, it is evident that TVP significantly influence the model's performance across various branches of the dataset. \textbf{Listening-before-Segmentation.} From Table~\ref{tab:ablation2}, we can observe that the performance of the model is significantly improved by the proposed modules. 
Among these modules, the CMPC module contributes the most significant improvement to the model's performance. This outcome is expected, as the CMPC module conducts fine-grained cross-modal alignment directly on patch tokens, which is most intrinsically related to the quality of pseudo-label generation.

\begin{figure}[tbp]
	\centering
	\includegraphics[width=0.9\linewidth]{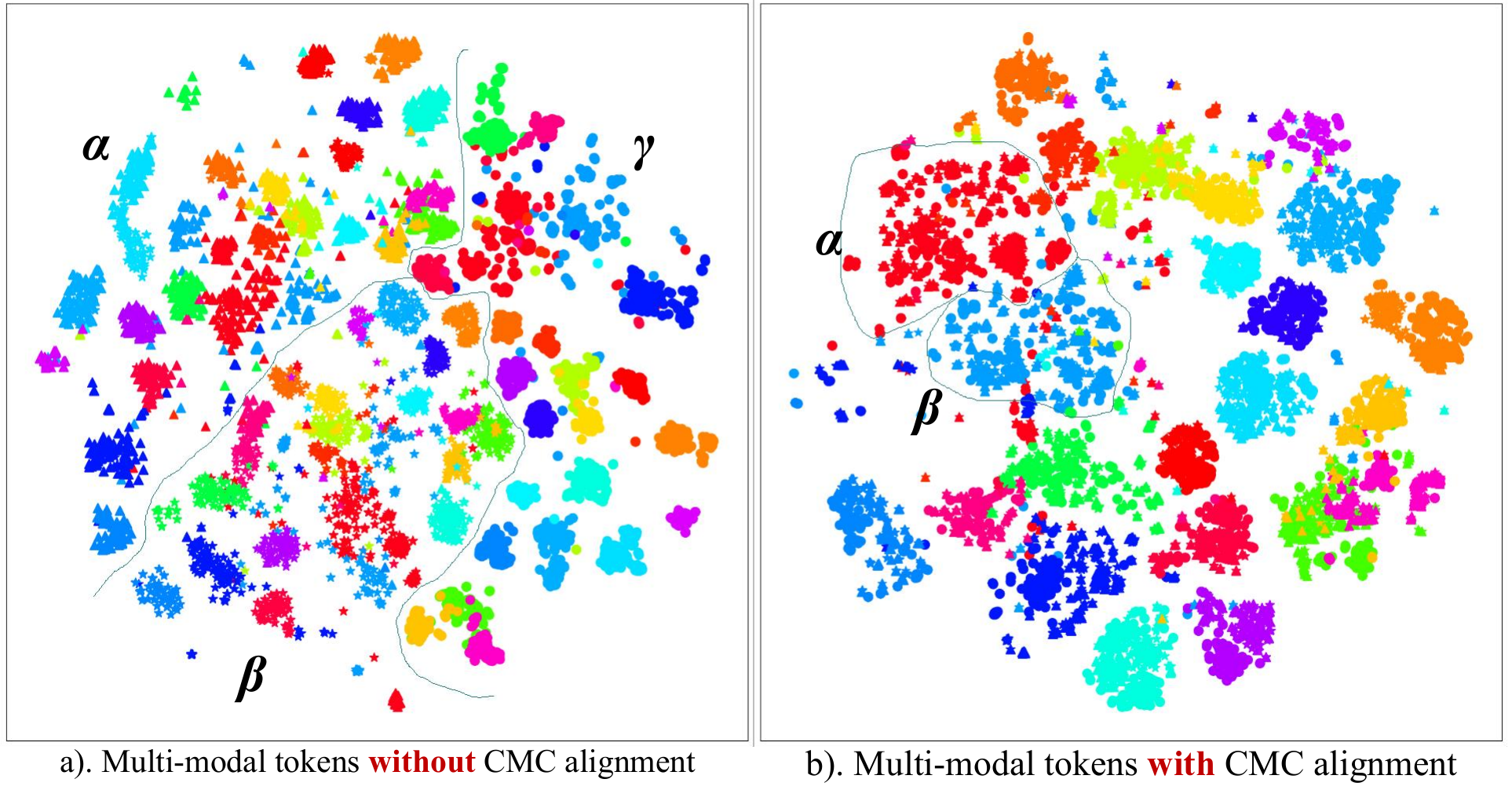}
	\caption{t-SNE of three token types with and without CMC. Points are samples; shapes indicate token types; colors indicate categories. \textquotesingle$\blacktriangle$\textquotesingle, \textquotesingle$\star$\textquotesingle, and \textquotesingle\textbullet\textquotesingle{} denote visual classification tokens ($v_{cls}$), visual semantic tokens ($v_{sem}$), and audio semantic tokens ($a_{sem}$), respectively.}
		\label{fig:t_SNE_c}
\end{figure}
\subsection{Feature Visualization}

We employ t-SNE~\cite{tsne} to reduce the dimensionality of AVSS sample features to 2D and visualize them. 
From figure~\ref{fig:t_SNE_c}a, it can be observed that \textquotesingle $\blacktriangle$\textquotesingle{} ($v_{cls}$), \textquotesingle$\star$\textquotesingle{} points ($v_{sem}$), and \textquotesingle\textbullet\textquotesingle{} points ($a_{sem}$) form clusters in the top-left $ \alpha $ region, bottom-left $ \beta $ region and the right $ \gamma $ region, respectively, indicating a misalignment between features of different modalities. In contrast, Figure~\ref{fig:t_SNE_c}b shows that after supervised contrastive learning, features from different modalities of the same category cluster together, with most features grouped according to their respective categories. \textbf{This comparison highlights the effectiveness of the CMC module.}

\begin{figure}[tbp]
	\centering
	\includegraphics[width=0.98\linewidth]{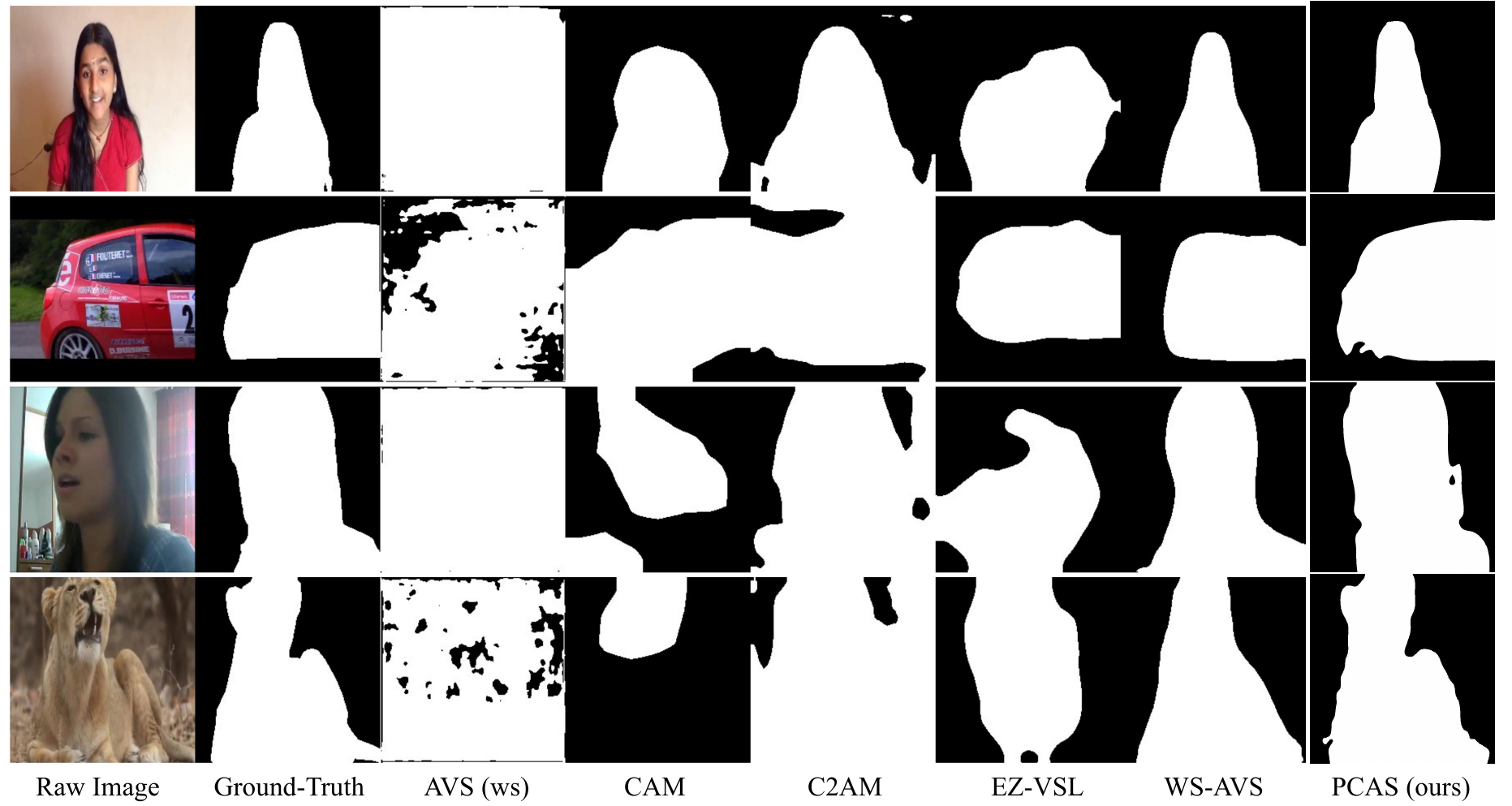}
	\caption{The comparison of output cases between PCAS and other weak supervision baselines on the AVS-S4 test dataset.}
		\label{fig:avs_case}
\end{figure}
\begin{figure}[tbp]
	\centering
	\includegraphics[width=\linewidth]{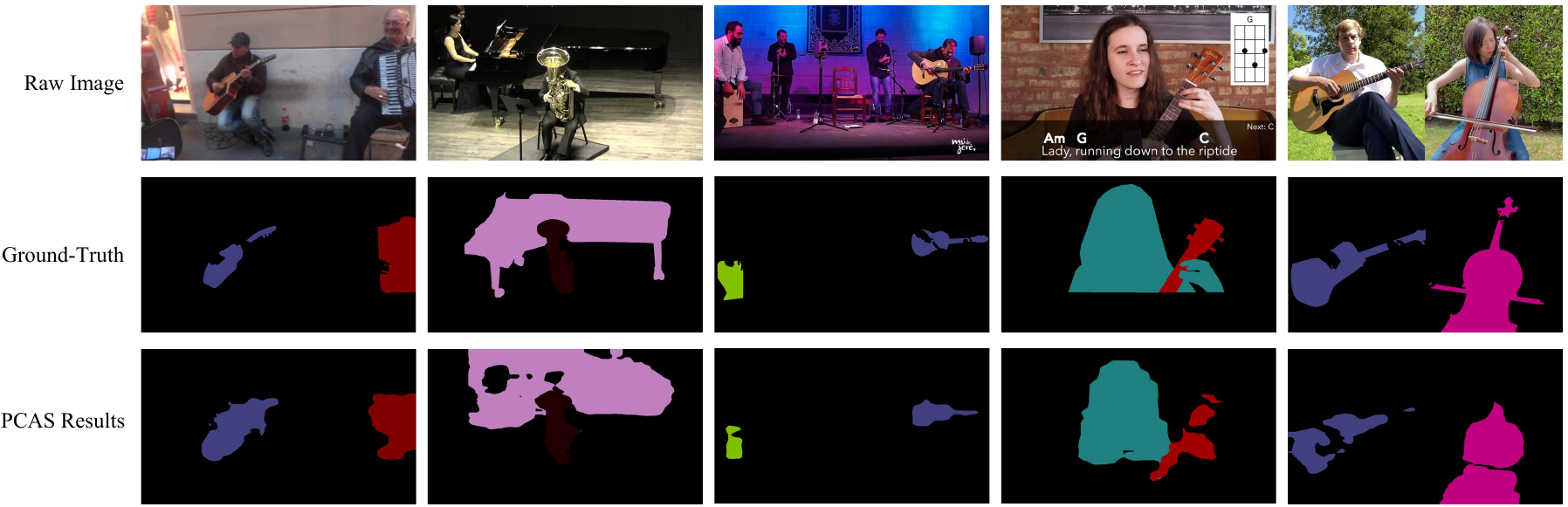}
	\caption{PCAS output cases on the AVSS test set. No weakly supervised baselines exist for comparison and shown.}
		\label{fig:avss_case}
\end{figure}

\subsection{Case Study}

As shown in Figure~\ref{fig:avs_case}, our method produces more accurate and detailed results than prior work. It better preserves fine semantic edges (e.g., shoulder contours and the cheetah’s ears and legs) that baselines miss, demonstrating the effectiveness of our proposed approach in achieving fine-grained semantic alignment across complex visual elements.. Figure~\ref{fig:avss_case} further shows that PCAS maps distinct audio categories to corresponding image regions on the AVSS dataset.

\section{Conclusion}
In this work, we introduce WSAVSS to reduce the costly pixel-level annotations required by AVSS. This is the first weakly supervised formulation of AVSS. Our PCAS method achieves performance comparable to fully supervised models on the challenging AVS-Semantic dataset.

\section{Acknowledgements}

This work is supported by the grants from the National Natural Science Foundation of China (No. 62376130 and U21B2009).

\bibliographystyle{IEEEbib}
\bibliography{references}

\end{document}